# Interfacial Layers between Ion and Water Detected by Terahertz Spectroscopy

Abhishek K. Singh, Luan C. Doan, Djamila Lou, Chengyuan Wen, Nguyen Q. Vinh*

Department of Physics and Center for Soft Matter and Biological Physics, Virginia Tech, Blacksburg, Virginia 24061, USA.

Corresponding author: *Email address: vinh@vt.edu; Phone: 540-231-3158

## Abstract

Dynamic fluctuations in hydrogen-bond network of water occur from femto- to nano-second timescale and provides insights into structural/dynamical aspects of water at ion-water interfaces. Employing terahertz spectroscopy assisted with molecular dynamics simulations, we study aqueous chloride solutions of five monovalent cations, namely, Li, Na, K, Rb and Cs. We show that ions modify the behavior of surrounding water molecules and form interfacial layers of water around them with physical properties distinct from that of bulk water. Small cations with high charge densities influence the kinetics of water well beyond the first solvation shell. At terahertz frequencies, we observe an emergence of fast relaxation processes of water with their magnitude following the ionic order Cs>Rb>K>Na>Li, revealing an enhanced population density of weakly coordinated water at ion-water interface. The results shed light on the structure breaking tendency of monovalent cations and provide insights into the properties of ionic solutions at the molecular level.

**Introduction**

Understanding the properties of aqueous salt solutions is critical for many branches of fundamental research and technology including the modeling of biological processes,[1-3] process design in the chemical industry,[4] drug design in the pharmaceutical industry,[5] electrolyte performance in fuel cells and batteries.[6] A large number of active experimental and theoretical work focuses on the behavior of neutral and charged surface immersed in electrolyte solutions.[7-11] However, many questions remain unanswered and our understanding of the structure and dynamics of water in ionic solutions at molecular level is still far from complete. Specifically, the behavior of water at interfaces of ion-water is complicated. The Coulombic potential of an ion modifies the behavior of surrounding water molecules, and forms interfacial layers of water around them with physical properties distinct from that of bulk water. The structural rearrangements of water molecules in these intermediate layers at ion-water interface modify the macroscopic properties of ionic solutions.[12-16] No consensus prevails so far regarding the length scales of electrostatic effects of ions in solution. A variety of experimental techniques have reported that ions affect primarily the dynamics and structure of the first solvation shell of water molecules which have direct contact with ions, and thus, endorsing local effects of ions into water. This includes optical Kerr-effect spectroscopy,[17] terahertz Kerr-effect spectroscopy,[10, 18, 19] two-dimensional Raman–terahertz spectroscopy,[20] femtosecond time-resolved infrared vibrational spectroscopy,[12, 13] terahertz spectroscopy,[14, 21-23] NMR,[24] dielectric spectroscopy,[25-31] and molecular dynamics simulations,[15, 32-34] that have significantly improved our understanding of aqueous salt solutions. However, long range effects of ions on water structure have also been witnessed.[14, 35, 36] Dielectric spectroscopy has shown mixed results with evidence that in certain cases the bound water molecules extend beyond the first solvation shell of ions.[21, 26-30, 37] Thus, detecting the structural dynamics of water at interfacial layers between ion and water as well as identifying their functions remains a significant challenge.

Ions in aqueous solution significantly alter the structure and dynamics of water molecules from their transient tetrahedral configuration.[7, 8, 12-14] Water molecules in the immediate neighborhood of ions experience a strong local electrostatic field, which significantly reduces the net polarization of water at ion-water interface. A strong charge density of a small ion causes a rearrangement of dipolar water molecules, forming hydration shells, whereas large ions interact weakly with water molecules. Moving ions orient adjacent water molecules in the opposite direction to that they would rotate,[38] further reducing the polarization of water. Thus, ions strongly influence the dynamics of water molecules beyond their hydration shells. Based on their influence beyond their hydration shells, ions have often been classified as structure making and structure breaking, or "*kosmotropic*" and "*chaotropic*", respectively, following the Hofmeister series of ions.[7] The dynamics of bulk water span a large range of frequencies from gigahertz to terahertz. Exploration of dynamics of water in this range will help us to understand the ion-water interaction. In this context, it is interesting to look into the interfacial layers (i.e., hydration shells and beyond) of water molecules around monovalent Hofmeister cations.

The dielectric response of aqueous salt solutions provides valuable information on the structure as well as dynamics of water and solvent in the solutions. The dielectric spectroscopy in the megahertz to gigahertz frequencies of ionic solutions has earlier been used to access the dynamics of hydrated ion-pairs and hydration shells.[25, 26, 28, 30, 37] There are only a few studies that provide detailed insights into the terahertz dielectric response of the aqueous salt solutions.[12, 14, 21, 23] Beyond the hydration shells, valuable information intrinsic to the effect of ions on water structure can be envisaged from fast relaxation processes of water in the aqueous salt solutions at terahertz frequencies. Such studies typically were overlooked because of the strong absorbance of water at these frequencies. Here, we study a series of aqueous salt solutions by means of an extended megahertz to terahertz dielectric spectroscopy assisted with molecular dynamics simulations. Measurements on chloride salts of five monovalent cations, namely, $Li^+$, $Na^+$, $K^+$, $Rb^+$, and $Cs^+$, with increasing size (or decreasing charge density) have been performed (Table 1). Based on the collective results over a series of five monovalent cations, we show that the order of increase in the fast water dielectric strengths observed at the terahertz frequencies is proportional to the decrease in orientational correlation among water molecules.

**Methods**

**Materials.** Dielectric spectroscopy measurements have been performed on aqueous solutions of alkali chlorides with five monovalent cations, namely, LiCl, NaCl, KCl, RbCl, and CsCl at room temperature (25 °C). The salts were obtained from Sigma Aldrich, USA. Dielectric response of four solutions with concentrations of 0.25, 0.50, 1.0, and 2.0 M were measured for each salt. The aqueous solutions were prepared by dissolving the salts into milli-Q water and stirring the solution for a few hours, and followed by equilibration at room temperature.

**Megahertz to terahertz dielectric spectroscopy.** To explore the effects of ions on aqueous solutions, we have employed a frequency-domain terahertz dielectric spectrometer covering a frequency range from 500 MHz to 1.12 THz (0.017 – 37.36 cm$^{-1}$).[14, 39-44] The spectrometer consists of a vector network analyzer (Agilent, N5225A PNA) and terahertz frequency extenders from Virginia Diodes. Frequency extenders from Virginia Diodes, Inc. (Charlottesville, VA), have been used to generate terahertz waves. Eight different rectangular-waveguide (WR) modules are employed to cover a frequency range of 9 GHz – 1.12 THz as described in previous works.[39] To characterize the low frequency dielectric response from 500 MHz to 50 GHz, a dielectric probe (HP 85070E) has been employed. At the terahertz frequencies, a variable-path length sample cell for the dielectric measurements has been used. Two parallel windows are installed inside the sample cell, with once fixed and other in mobile position with submicron (~80 nm) precision for changing thickness. To control the temperature, the sample cell was cooled down with Peltier coolers (Custom Thermoelectric, 27115L31-03CK), and heated up with high power resistors embedded in the aluminum sample cell. The temperature is controlled with accuracy of ± 0.02 ºC, using a Lakeshore 336 temperature controller. Additional information regarding experimental setup can be accessed in earlier reports.[14, 39-46]

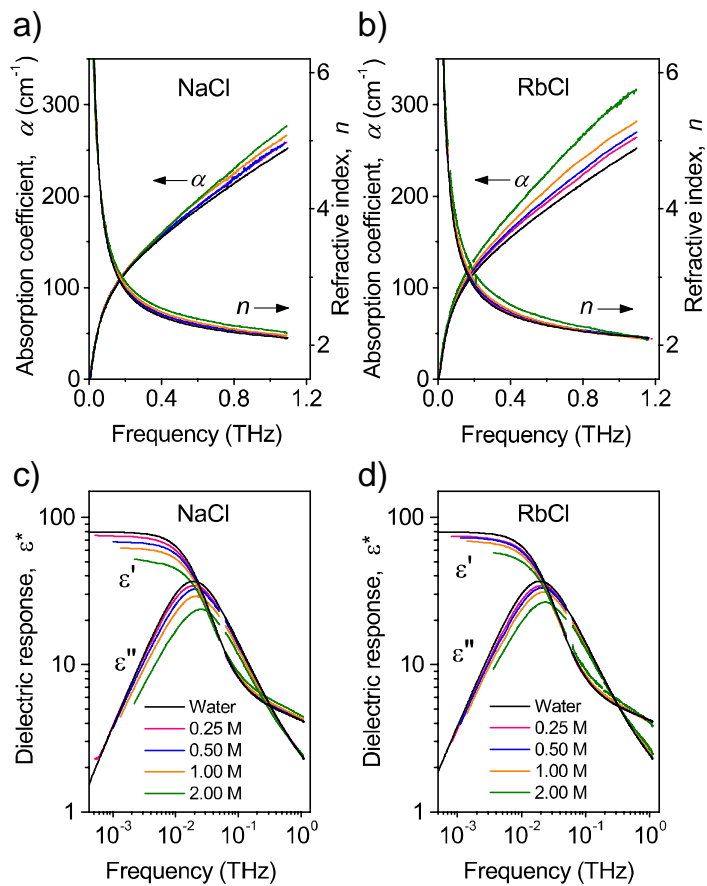

**FIG. 1. Dielectric response of aqueous salt solutions at megahertz to terahertz frequencies revealing the interaction between ion-water.** (a) The absorption coefficients, $\alpha(v)$, and (b) refractive indices, $n(v)$, for water and aqueous NaCl, RbCl solutions over a large range of concentrations (0.25, 0.50, 1.0 and 2.0 M). The corresponding dielectric response, $\varepsilon^*(v)$, composed of real, $\varepsilon'(v)$, and imaginary parts, $\varepsilon''(v)$, were obtained from the absorption coefficient and refractive index for (c) NaCl and (d) RbCl, respectively.

**Molecular Dynamics Simulations.** Molecular dynamics simulations are carried out to access molecular level insights and complement our experimental results of aqueous salts solutions. All the atomistic simulations are performed by means of the Large-scale Atomic/Molecular Massively Parallel Simulator (LAMMPS).[47] The SPC/E water model is used as it provides relaxation times for bulk water compatible to that obtained from the dielectric measurements. Details of the simulations are provided in supplementary material.

## Results

**Dielectric spectroscopy.** The dynamics of water and ions in aqueous salt solutions are complex in nature, encompassing several inter/intra-molecular relaxations. In the megahertz to gigahertz frequencies, the dynamic processes typically involve the tumbling of ion-pairs, orientational modes of water molecules in solvation shells of ions, and in bulk water.[8, 14, 25, 28, 39-43, 48] At terahertz frequencies, the structural rearrangement due to breaking and reforming hydrogen bonds of water molecules as well as the orientation of single-water molecules becomes prominent, originating from the interaction between ions and water molecules in the aqueous solutions.[14, 49-52] The complex refractive index is conventionally represented as a function of frequency, $v$, in the form,

$$n^*(v) = n(v) + i\kappa(v) \tag{1a}$$

where $n(v)$ is refraction index, $\kappa(v) = c\alpha(v)/(4\pi v)$ is extinction coefficient of the solution with $c$ being speed of light, and $\alpha(v)$ is the absorption coefficient. The absorption coefficient and refractive index are shown in Figs. 1a and 1b for pure water along with 0.25, 0.50, 1.0, and 2.0 M NaCl and RbCl solutions, respectively. The similar results for LiCl, KCl, and CsCl solutions are provided in supplementary material, Figs S1a, S1b, and S1c. The solvation of salt species into water significantly changes the absorption coefficient as well as refractive index. A significant increase in the terahertz absorbance with respect to water has been observed in salt solutions containing cations with large ionic size, such as $K^+$, $Rb^+$, and $Cs^+$, as compared to smaller ones, i.e., $Li^+$ and $Na^+$, reflecting the diverse nature of ion effects on aqueous solutions.

The real, $\varepsilon'(v)$, and imaginary, $\varepsilon''(v)$, parts of the complex dielectric response can be calculated from the absorption coefficient and refractive index using the following equation,

$$\varepsilon^*_{sol}(v) = \varepsilon'(v) + i\varepsilon''(v)$$
$$= \left(n^2(v) - \left(\frac{c\alpha(v)}{4\pi v}\right)^2\right) + i\left(\frac{2n(v)c\alpha(v)}{4\pi v} - \frac{\sigma}{2\pi v\varepsilon_0}\right), \tag{1b}$$

where $\sigma/(2\pi v\varepsilon_0)$ is the Ohmic loss due to the electrical conductivity, $\sigma$, of the salt solution (Fig. S2), and $\varepsilon_0$ is the permittivity of the vacuum. The real and imaginary parts of the dielectric response are shown in Figs. 1c and 1d, for NaCl and RbCl solutions, respectively. The dielectric spectra for LiCl, KCl and CsCl solutions are presented in supplementary material, Figs. S1d, S1e, and S1f.

The interaction between ions and water molecules strongly depends on the charge density of ions.[53] Small monovalent ions with a high charge density strongly interact with surrounding water molecules, whereas such interactions are relatively weak for large ions with a low charge density. To understand relaxational modes of water molecules in ionic solutions, the dielectric response in the megahertz to gigahertz frequency range has extensively been studied.[14, 26, 28, 54] The technique provides crucial information about the dynamics of hydration water. The dynamics of hydration water having direct contacts with ions depends strongly on the charge density of ions, and varies from ~10 to ~55 ps. As an example, the dynamics of hydration water in the 0.25 M LiCl solution of ~55 ps (~2.8 GHz) has been observed (Fig. S3). This value is of the same order as that reported in the literature.[25, 55] The dynamics of hydration water around large monovalent ions such as $K^+$, $Rb^+$, $Cs^+$ occurs in the gigahertz to terahertz frequencies, and is more or less similar with the orientational dynamics of pure water.[25, 27, 56] Thus, we focus on the analysis of the dielectric response at the gigahertz to terahertz frequencies.

The presence of solutes with charged surfaces changes the gigahertz to terahertz frequency dielectric response of water.[2, 3, 14, 43, 57] Heterogeneous dynamics of water has been reported around ions, proteins, among other species. Debye model composed of several Debye components has proven to be useful to address the dynamic heterogeneity in a broad range of aqueous solutions.[29, 58] In that view, the dielectric response of aqueous salt solutions in this frequency range can adequately be analyzed using a Debye model containing three individual components in the form,[14, 50, 51]

$$\varepsilon^*_{\text{sol}}(\nu) = \varepsilon_\infty + \frac{\varepsilon_S - \varepsilon_1}{1 + i2\pi\nu\tau_D} + \frac{\varepsilon_1 - \varepsilon_2}{1 + i2\pi\nu\tau_2} + \frac{\varepsilon_2 - \varepsilon_\infty}{1 + i2\pi\nu\tau_3}, \qquad (2)$$

where $\tau_D$ is the relaxation time corresponding to the collective orientational dynamics of bulk water in the solution. The relaxation times, $\tau_2$ and $\tau_3$, are the fast relaxation processes, including the orientation of single-water molecules and the structural rearrangement due to breaking and reforming hydrogen bonds around solvated ions, respectively. The numerators associated with each term, $\Delta\varepsilon_D = \varepsilon_S - \varepsilon_1$, $\Delta\varepsilon_2 = \varepsilon_1 - \varepsilon_2$ and $\Delta\varepsilon_3 = \varepsilon_2 - \varepsilon_\infty$, correspond to the dielectric strengths of the corresponding processes. $\varepsilon_\infty$ is the dielectric contribution from all polarization modes at frequencies much higher than the range probed in our dielectric measurements. And $\varepsilon_S$ is the static permittivity given by $\varepsilon_S = \varepsilon_\infty + \Delta\varepsilon_D + \Delta\varepsilon_2 + \Delta\varepsilon_3$. Solvation of salts into water significantly alters the hydrogen-bond network in bulk water and can been recognized from the dielectric response of the aqueous salt solutions. As shown in Figs. 1 and S1, the absorption coefficients change significantly as the salt concentration increases. The peak value of the dielectric loss reduces with increasing salt for all solutions.

To quantify individual contributions from different relaxation processes, the dielectric response of aqueous salt solutions is analyzed using Eq. 2. An example of such fits of the dielectric response to Eq. 2 is shown in Fig. 2, for pure water, 0.25 and 2.0 M RbCl solutions. The relaxation times extracted for pure water are $\tau_D = 8.27 \pm 0.20$ ps, $\tau_2 = 1.1 \pm 0.3$ ps, and $\tau_3 = 0.16 \pm 0.05$ ps.[14, 50-52] Similar relaxation times are observed within experimental errors for different solutions of concentrations and salts. The relaxation times and dielectric strengths of water in the salt solutions as a function of salt concentration have been presented in Figs. 3 and S4. The dielectric strength of bulk water in the salt solutions reduces, while the dielectric strengths of fast relaxation processes increase.

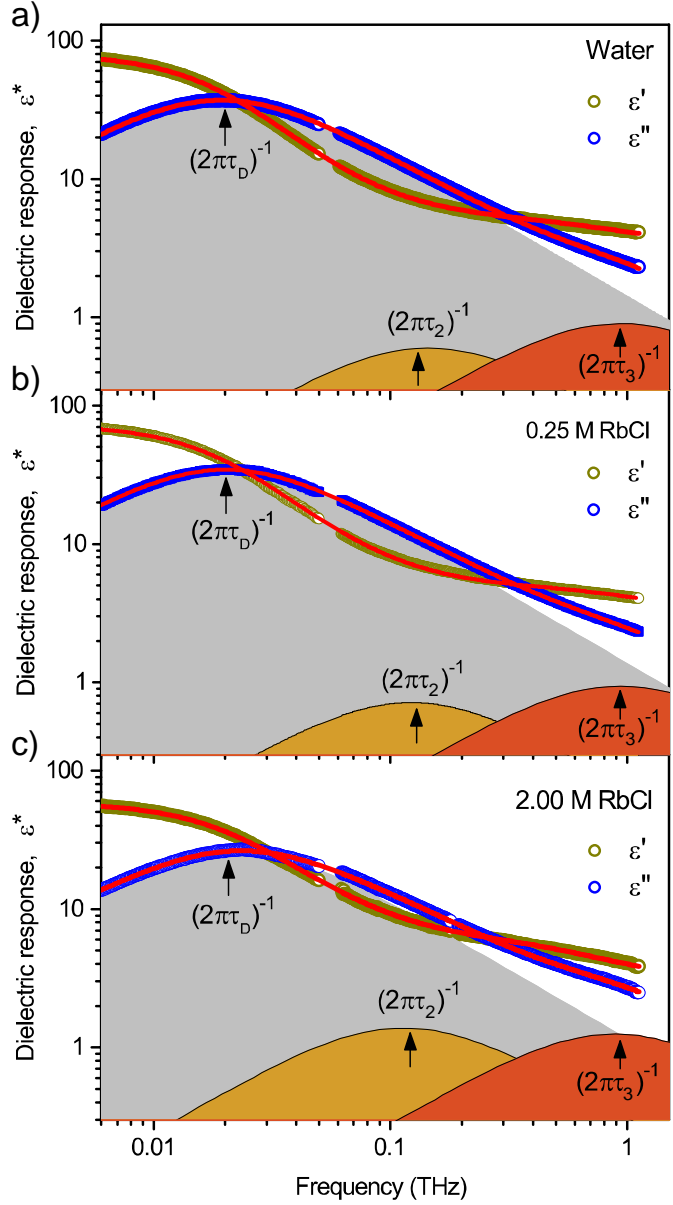

**FIG 2. Dielectric response, $\varepsilon^*(\nu)$, at gigahertz to terahertz frequencies.** (a) water, (b) 0.25 M RbCl, and (c) 2.0 M RbCl solutions providing relaxational dynamics of water molecules in the solutions. The red curves are fits of the real and imaginary components of the dielectric response of the solutions to Eq. 2. Individual contributions represent for the collective orientational dynamics among water molecules (bulk water), $\tau_D$ (shaded grey area), and fast relaxation modes including to $\tau_2$ (shaded yellow area) and $\tau_3$ (shaded rust area). An increase in the amplitude of the fast relaxational dynamics of water in the salt solutions has been observed.

**Molecular Dynamics Simulations.** A microscopic view of the spatio-temporal behavior of water molecules in hydration shells around an ion can be accessed using molecular dynamics (MD) simulations.

With the help of MD simulations, the probability distribution of water molecules with respect to the ion and their dynamics in aqueous salt solutions can be visualized. The arrangement of water molecules with respect to the ion depends on the nature of the charge as well as their charge densities. For instance, cations with positively charged surface directly interact with oxygen of water molecules, whereas anions with negatively charged surface have a contact with hydrogen atoms of water molecules.

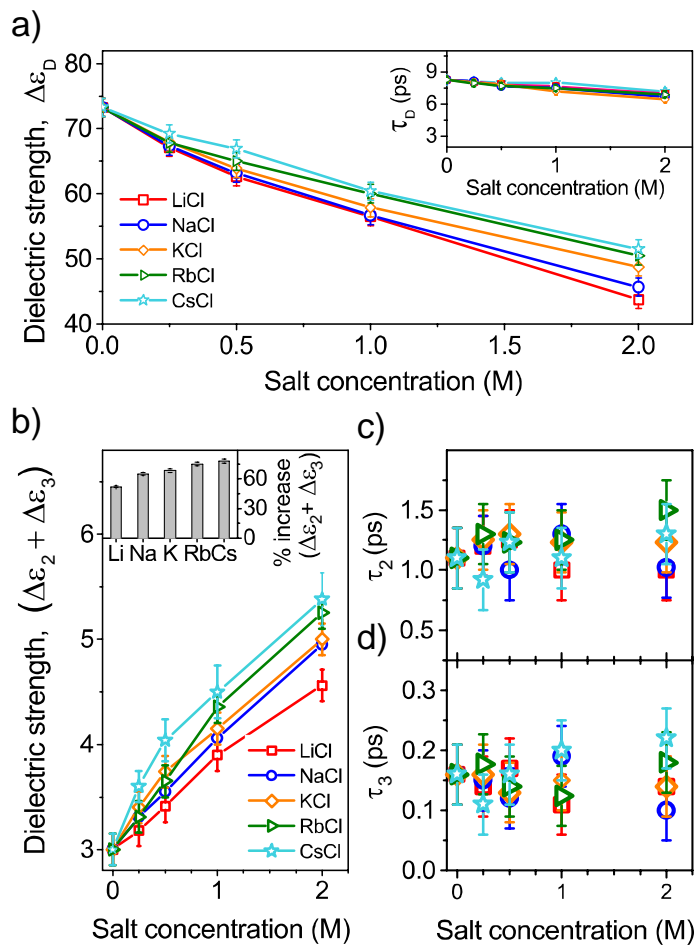

**FIG. 3. The variation of the dielectric strength and dynamics of relaxation processes of water in salt solutions with increasing salt concentration providing insights into their origins.** (a) The dielectric strength of the $\tau_D$ relaxation mode reduces monotonically when the salt concentration increases for all the salt solutions. The corresponding relaxation times are shown in the inset. (b) The total dielectric strengths of fast relaxation dynamics increase with increasing salt concentration. As shown in the inset, the net increase in the total dielectric strengths of fast relaxation modes from pure water to 2.0 M salt concentration is more pronounced in case of large cations. (c) The relaxation times of the fast relaxation processes plotted against with salt concentration are similar with those in bulk water.

In order to probe the interaction of water with cations, we estimate the probability distribution of oxygen of water molecules with respect to ions, $X^+ – O$, where X represents a cation and O is the oxygen of water (Fig. 4a, for 0.5 M salt solutions). The first peak in cation-water radial distribution functions, $RDF_{X-O}(r)$, indicates the occurrence of the first hydration shell with respect to the X cation. The position and amplitude of the first peaks agree closely with previous results.[56] The position of the peak occurs at a shorter distance for small cations, such as $Li^+$ and $Na^+$, indicating that water molecules in the hydration shells are strongly bound to these cations with their charge density confined to a narrow volume. Whereas the first peak in radial distribution functions of bigger cations with their charge distributed over a larger volume, such as $K^+$, $Rb^+$ and $Cs^+$, appears at a longer distance with respect to the cations, revealing that hydration water molecules within the first solvation shell of these cations have a relatively weaker interaction. In addition, the first peaks in these functions of bigger cations are broader, indicating that water molecules are rather diffused in the space around the surface of large cations with lower surface charge densities. The interaction between hydrogen atoms of water with anion ($Cl^-$) also has been characterized with the radial distribution function (Fig. 4a, inset). The first peak of the probability distribution of hydrogen atoms of water with respect to $Cl^-$ ($Cl^- – H$) emerges at a long distance of ~ 3.10 Å. This peak appears at almost the same position of the first peak in cesium-oxygen, $Cs^+ – O$, (~ 3.30 Å), and this distance is similar with the distance between water molecules in water-water radial distribution function (~ 3.30 Å),[56] indicating a minor difference between the interactions of water with $Cl^-$, $Cs^+$, and water.

**Table 1:** Molecular Dynamics simulations for monovalent ions in aqueous solutions. The radius of ions, the coordination number ($CN$), the mobility ($\mu_i$ in $cm^2V^{-1}s^{-1} \times 10^{-4}$)[59] of aqueous ions at 25 ºC, and the Jones-Dole Viscosity ($B$), coefficient[60] were obtained from literature. Positions of minima of radial distribution functions of cation-oxygen and anion-hydrogen are specified. The orientational correlation functions, $C(t)$, are fitted to a three-exponential component function with weighted factors ($A$, $B$ and $C$) and relaxation times ($\tau_A$, $\tau_B$, and $\tau_C$).

| Ions | Ionic radius (pm) | CN | $\mu_i$ | B | First minimum (Å) | Second minimum (Å) | A | B | C | $\tau_A$ ± 1 (ps) | $\tau_B$ ± 2 (ps) | $\tau_C$ ± 30 (ps) |
|---|---|---|---|---|---|---|---|---|---|---|---|---|
| $Li^+$ | 60[61] | 4[61,62] | 4.01 | 0.147 | 2.5 | 5.0 | -- | 0.90 | 0.10 | -- | 32.0 | 180 |
| $Na^+$ | 102[61] | 5[61], 5.7[32] | 5.19 | 0.086 | 3.1 | 5.4 | -- | 0.88 | 0.12 | -- | 25.0 | 160 |
| $K^+$ | 146[61] | 7[32], 8.3[63] | 7.62 | -0.007 | 3.7 | 5.7 | 0.51 | 0.47 | 0.02 | 5.2 | 12.0 | 150 |
| $Rb^+$ | 164[61] | 8[61,64] | 7.92 | -0.029 | 3.8 | 5.9 | 0.56 | 0.42 | 0.02 | 5.1 | 11.5 | 150 |
| $Cs^+$ | 173[61] | 8[61,62], 10[64] | 7.96 | -0.045 | 4.4 | 5.9 | 0.64 | 0.32 | 0.03 | 5.9 | 10.5 | 140 |
| $Cl^-$ | 182[65] | 6[8,66], 7.4[56] | -- | -- | 3.8 | 6.1 | 0.61 | 0.36 | 0.03 | 5.0 | 8.8 | -- |

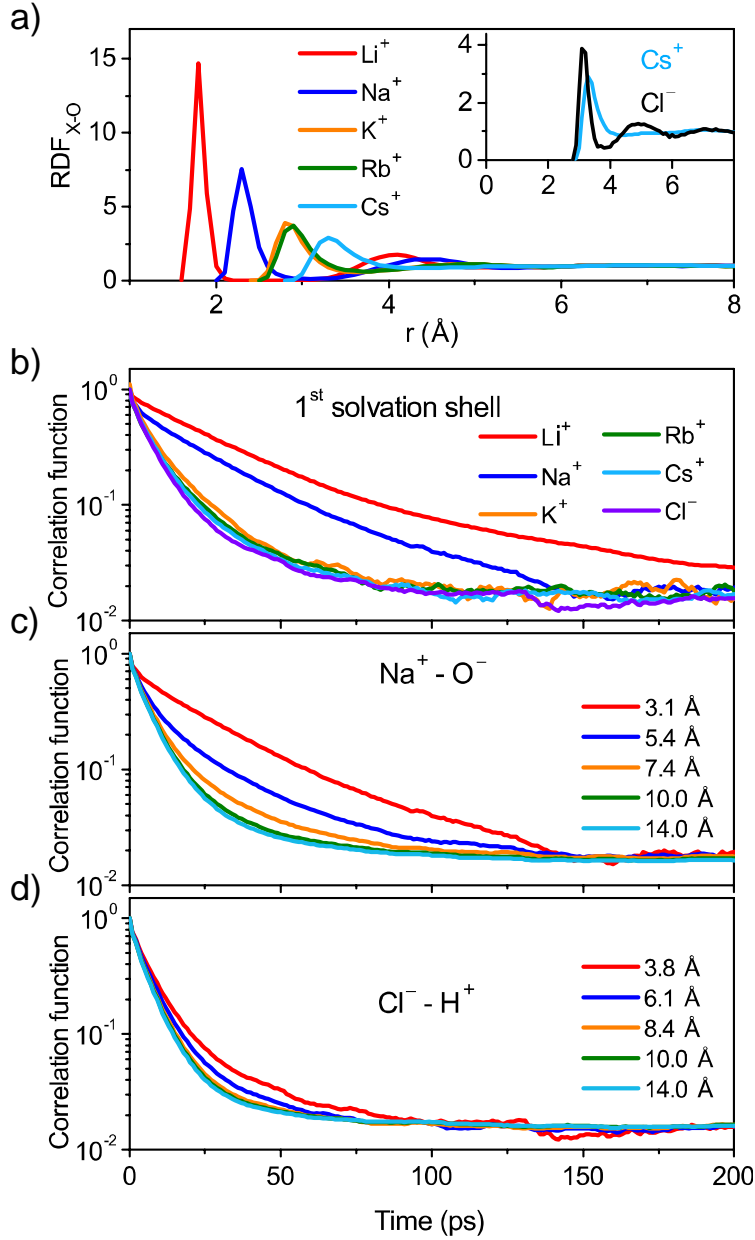

**FIG. 4. Microscopic view of solvation dynamics for aqueous salt solutions from molecular dynamics simulations.** (a) Radial distribution functions, $RDF_{X-O}(r)$, are shown as a function of the distance between the oxygen in water molecules and the cation, for $Li^+$ (red), $Na^+$ (blue), $K^+$ (orange), $Rb^+$ (dark yellow), and $Cs^+$ (light blue). Inset shows the $RDF_{X-O}(r)$ for $Cl^-$ as a function of the distance of hydrogen atoms in water molecules with respect to the anion. (b) Cation-water correlation functions, $C(t)_{X-O}$, reveal the dynamics of water in the first hydration shell for each cation. The anion-water correlation function, $C(t)_{Cl^--H}$, of $Cl^-$- water is shown for comparison. The cation-water correlation functions, $C(t)_{X-O}$, are calculated at different distances from the cation and the oxygen in water molecules for (c) $Na^+$, (d) $Cl^-$.

The dynamics of water molecules in the first solvation shell can be extracted from the orientational correlation function (OCF), $C(t)$, as shown in Fig. 4b. The OCFs show three exponential decay characteristics and can be analyzed in the form,

$$C(t) = A \exp(t/\tau_A) + B \exp(t/\tau_B) + C \exp(t/\tau_C),  \quad (3)$$

where $A$, $B$ and $C$ are the weighted factors; and $\tau_A$, $\tau_B$ and $\tau_C$ are relaxation times for the three orientational processes. Examples of fits of the OCF data to Eq. 3 are provided in supplementary material, Fig. S5. The weighted factors and relaxation times of water molecules in the first solvation shell of six ions are included in Table 1. The fast component, $\tau_A$, has the orientational relaxation time of $5.3 \pm 1$ ps, which is similar to the dynamics of bulk water characterized by the dielectric spectroscopy.[14, 50, 51] For pure water, the orientational relaxation time of water molecules has been determined using the SPC/E water model to be $5.2 \pm 0.6$ ps.[41, 42, 45, 46, 55, 56] Thus, the orientation relaxation time, $\tau_A$, corresponds to the relaxational dynamics of bulk water in the salt solutions. A relatively slower component characterized by the relaxation time, $\tau_B$, is observed, corresponding to the orientational dynamics of water molecules in the first solvation shell of the ion. This value strongly depends on the charge density of the cation (Table 1). An additional component with a weak contribution to the total spectrum has also been identified with the relaxational time ranging from 140 to 180 ps, which appears to be the tumbling motion of ion-pairs that has previously been observed in monovalent cations.[25, 26] The relaxation dynamics of water molecules in the first hydration shell of $Li^+$ and $Na^+$ is significantly slower than those of $Rb^+$ and $Cs^+$ (Fig. 4b). Relaxation dynamics of water in the first hydration shell of $Li^+$ and $Na^+$ show two long exponential decay components with no contribution from bulk water relaxation (Table 1), demonstrating a strong interaction of $Li^+$ and $Na^+$ with surrounding water molecules. However, in case of $K^+$, $Rb^+$ and $Cs^+$, all the three dynamical processes of water molecules are present. Water molecules in the first solvation shell of the large ions having a weak cation-water interaction easily exchange with bulk water, and therefore the contributions of bulk water in the first solvation shell appears in the OCFs. The relaxation times of water molecules in the first solvation shell of ions (i.e., slow water) are 32, 25, 12, 11.5, 10.5 and 8.8 ps for $Li^+$, $Na^+$, $K^+$, $Rb^+$, $Cs^+$, and $Cl^-$, respectively (Table 1). Clearly, the orientational dynamics of water molecules in the first hydration shell of small cations is kinetically more retarded as compared to that of the large ions. The orientational dynamics of water molecules interacting with $Cl^-$ is the least affected one among all the ions and is nearly indistinguishable from that of bulk water (Table 1).

The length-scale of the interaction between an ion and water molecules can be estimated from the OCF, $C(t)_{X-O}$, of water molecules around ions. The OCFs provide information of the dynamics of water molecules around an ion as well as qualitative length-scales, i.e., how far out into the solution the influence of an ion can reach. To explore the dynamics of water around an ion, the correlation functions are determined between the surface of the ion and successive solvation shells (i.e., local minima in the radial distribution function). The OCFs of water around ions as a function of distance from the ion have been analyzed until 14 Å for $Na^+$, $Cl^-$, $Li^+$, $K^+$, $Rb^+$ and $Cs^+$ (Figs. 4c, 4d and S6). The length-scale of the ion-water interaction can be estimated from the ion to the position, at which the OCF becomes nearly invariant with increasing water thickness. Based on this criterion, the dynamics of water molecules is influenced until ~ 6.9 Å from $Li^+$. Beyond the distance, the OCF remains unchanged and is mostly governed by bulk water dynamics. For $Na^+$ and $K^+$ ions, the dynamics of water molecules is affected until about 5.4 Å and 3.7 Å, respectively. For the large cations, $Rb^+$ and $Cs^+$, only water molecules in the first solvation shell are somewhat affected. The observation has been confirmed by the gigahertz dielectric response, suggesting that small cations with high charge densities have more tendency to accumulate water molecules around them.

**Length-scale of cation-water interaction determined by the gigahertz spectroscopy.** The presence of salt in aqueous solutions significantly influences the dynamics of water molecules, which reflects in the dielectric response of ionic solutions. The dominant process characterized by the relaxation time $\tau_D$ in the dielectric response of aqueous salt solutions occurs at 19.25 GHz (~ 8.27 ps) attributes to the collective orientational motion of bulk water in the solution.[50, 51] The strong local electrical field of ions in the aqueous solution leads to a reduction in the dielectric response of water, which is referred to as "*depolarization*".[28, 48] As a result, the dielectric strength of the main relaxation process monotonically decreases with increasing salt concentration (Fig. 3a). The reduction of the dielectric response has consistently been observed in salt solutions and originates from accumulative effects that include the dilution of water, and the depolarization of water due to the presence of ions.[25, 26, 29, 30, 37, 48, 54] To correct for the dilution of water, we calculate the dielectric response originated from bulk water in the solution. If we assume that all water molecules in the solution participate in the bulk water process, their dielectric strength can be estimated based on the partial specific volume of water in the solution, which is the line labeled as "*ideal bulk water*" in Figs. 5a, 5b and S7. The depolarization can be separated into (*i*) the interaction of water molecules with a mobile ion in an electrical field (i.e., the *kinetic depolarization*), and (*ii*) the reduced orientation of water molecules in hydration shells of ions (i.e., the *static depolarization* or *hydration effect*). The kinetic depolarization arises from the reorientation of surrounding water molecules in opposition to the electrical field of ions, lowering the polarization of water molecular dipoles.

The Onsager – Hubbard continuum theory[67] predicts the kinetic depolarization for slip boundary conditions at the ion surface for a salt solution with concentration of $c_s$.[28, 48]

$$\Delta_{kd}\varepsilon(c_s) = \frac{2}{3}\left(\frac{\varepsilon_S(0)-\varepsilon_\infty(c_s)}{\varepsilon_S(0)}\right)\frac{\tau_D(0)}{\varepsilon_0}\sigma(c_s), \tag{4a}$$

where $\sigma(c_s)$ is the electrical conductivity of the aqueous solution (Fig. S2). The dark brown curves (in Figs. 5a, 5b and S7) represent the correction due to the contribution from the kinetic depolarization. The dielectric contribution of the static depolarization then can be extracted and used to estimate the number of slow water molecules per solvated ion, i.e., the "*hydration number*". The orientational dynamics of these water molecules appears at the lower frequency (Fig. S3). Deconvolution of the dielectric response at the megahertz to gigahertz frequencies reveals an emerging contribution from hydration water molecules around ions (Figs. 5a, 5b and S7). Such estimation has been carried out for a large number of aqueous salt solutions,[25-27] using the relationship,

$$N_{\text{hyd}} = \left(c_{\text{w}} - \frac{\Delta\varepsilon_{\text{w}} + \Delta_{\text{kd}}\varepsilon(c_s)}{\Delta\varepsilon_{\text{pure}}} c_{\text{pure}}\right)/c_s, \qquad (4b)$$

where $c_{\text{pure}} = 55.35$ M is the molar concentration of pure water, $c_{\text{w}}$ is the concentration of water in the solution, $\Delta\varepsilon_{\text{pure}}$ is the dielectric strength of pure water, $\Delta\varepsilon_{\text{w}} = \varepsilon_S(c_s) - \varepsilon_\infty(c_s)$ is the dielectric strength of bulk water in the salt solution with concentration $c_s$. Hydration numbers estimated based on modified Hubbard-Onsager theory[38] are also shown in Fig. S8.

The hydration number corresponds to the total number of water molecules that are kinetically retarded (i.e., slow water) due to presence of ions in the solution. The hydration numbers estimated using the above procedure for all salt solutions with different concentrations are shown in Fig. 5c. Several earlier investigations have pointed out that the dynamics of water molecules around chloride ions is largely indistinguishable from that of bulk water.[20, 68] Molecular level insights supporting this aspect are provided with the help of molecular dynamics simulations. Based on these results, the estimation of hydration water in salt solutions can be attributed to the cationic hydration shell only. The number of slow water molecules is reduced when the salt concentration increases (Fig. 5c). The decrease in the hydration number at high salt concentrations suggests the overlapping of hydration layers or the forming ion clusters in the salt solutions. Thus, the hydration number has been determined at lower concentrations of 0.25 or 0.5 M (Fig. 5d).

A comparison between the effective hydration number characterized from dielectric response and the number of water molecules immediately surrounding an ion (i.e., coordination number) sheds light on the effect of cations on the solution. The results for 0.25 M salt solutions (Fig. 5d) indicate that the small cations have a higher number of slow water (i.e., hydration number) molecules as compared to the large ones. An approximate analysis can be carried out by comparing the hydration number with the coordination number, *CN*, of cations included in Table 1. As per coordination number of Li$^+$ (Table 1), about four water molecules are accommodated within the first hydration shell. The hydration water (~ 8.5) extracted from the dielectric spectroscopy exceeds the first hydration shell. Thus, the total hydration water molecules extend to the second hydration shell around Li$^+$ cation. It is noteworthy that water molecules outside the hydration shells are also affected by the structural discrepancy between the neat water (i.e., bulk water) and the hydration water. The structural dynamics of these water molecules is modified as a result of ion-water interaction. In case of $Na^+$, hydration water molecules (~ 4.5) can be accommodated well within the first solvation layer, while total water molecules affected by the ion may extend to the second solvation shell. Large monovalent cations ($K^+$, $Rb^+$, and $Cs^+$) have a similar coordination number, i.e., about 8 (Table 1). However, the hydration water molecules bound to $K^+$ is ~2.2, whereas this value is close to zero for $Rb^+$ and $Cs^+$. Water molecules around large monovalent cations do not experience a strong electrostatic effect, however, they still are influenced by the electrical charge on the surface of the ions. Thus, the structural aspects of these water molecules are not the same as bulk water.

**Interfacial dynamics between bulk water and hydration water observed with the terahertz spectroscopy.** Terahertz spectroscopy provides valuable insights into the dynamics of water molecules beyond the hydration shells of ions. The dielectric response of salt solutions at terahertz frequencies is governed by two distinct orientational modes arising from weakly coordinated water molecules in the intermediate layer at the ion-water interface, including the reorientation of single-water molecules, $\tau_2$,[14, 50, 51, 69] and the structural rearrangement due to breaking and reforming hydrogen bonds around solvated ions, $\tau_3$.[14, 50-52, 70, 71] The single-water molecules have weak hydrogen bonds to their neighboring molecules, lacking at least three of their four potential hydrogen bonds.[14, 72, 73] Molecular dynamics simulation studies show that while most of the water molecules are in the approximately tetrahedral configuration at a time, a very small fraction, ~3 %, of them exist with one or lesser hydrogen bonds.[70, 74, 75] Moreover, due to structural rearrangement of hydrogen bonds around the solvated ions, the fastest dynamics emerges from the rapid disruption and reformation of individual hydrogen bonds,[14, 50, 51] which has earlier been suggested from simulations and quantum mechanical computations.[75] The timescales of these fast relaxation modes of are obtained from the dielectric response of pure water are $\tau_2 = 1.10 \pm 0.25$ ps and $\tau_3 = 0.16 \pm 0.05$ ps.[14, 50-52] The corresponding dielectric strengths are about 1.8 % and 2.1% with respect to the total dielectric response of bulk water.

The fast dynamical processes of water in aqueous salt solutions have been characterized by the terahertz dielectric response. The dielectric strengths of both fast components increase with increasing salt concentration (Fig. 3b), resulting from an increase in weakly coordinated water molecules in the intermediate layer between hydrated ions or hydration layers and bulk water.[14, 50, 51] The increment is more pronounced in large cations. An increase of about 52, 65, 68, 75, and 77 % in the total dielectric strengths of the fast relaxational modes in 2 M salt solutions has been observed for $Li^+$, $Na^+$, $K^+$, $Rb^+$, and $Cs^+$ cations, respectively, with respect to that of bulk water (Fig. 3b, inset). In view of structure breaking tendency of ions, the total dielectric strengths of the fast relaxation modes of water in alkali chloride solutions can be arranged in the order as, $Cs^+ > Rb^+ > K^+ > Na^+ > Li^+$. The observed trend infers in the first place that cations with a larger size give rise to a higher number of weakly coordinated water molecules in the solution. This behavior conforms well with the classical Hofmeister series, that advocates the structure breaking tendency of these cations to be in the order, $Cs^+ > Rb^+ > K^+ > Na^+ > Li^+$.[9, 53]

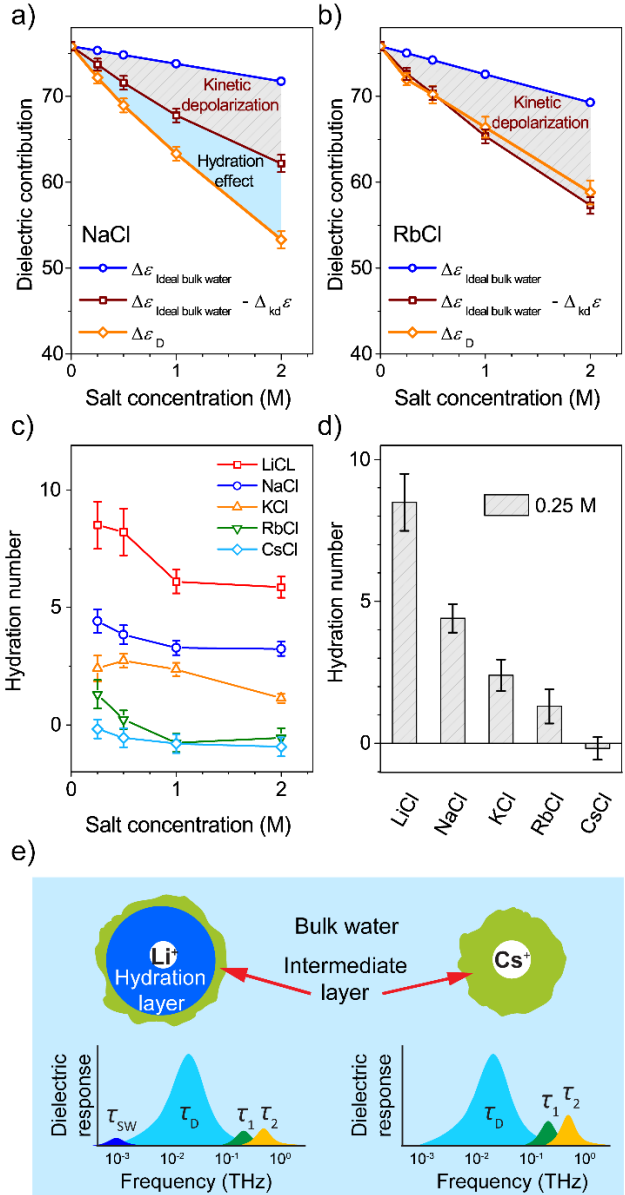

**FIG. 5. Hydration effect of ions in solution.** The dielectric strength of the collective orientational dynamics, $\tau_D$, was analyzed into kinetic and static depolarization processes for (a) NaCl and (b) RbCl solutions at different salt concentrations. (c) The hydration numbers are extracted for LiCl, NaCl, KCl, RbCl, and CsCl solutions. (d) At high salt concentrations, hydration shells are overlapped, thus, the hydration number has been determined at 0.25 M. (e) A schematic representation illustrates interfacial layers (hydration and intermediate) of water around Li and Cs ions along with relaxation modes of slow, fast, and bulk water relaxation processes.

**Discussion**

The dynamic response of liquid water extends from gigahertz to terahertz frequencies encompassing several dynamical features that are of inter- and intramolecular origin. In this work, we have analyzed the dielectric response from liquid water at 100 MHz – 1.12 THz using a three-Debye model and have identified three distinct dynamics with relaxation times of $\tau_D = 8.27 \pm 0.20$ ps, $\tau_2 = 1.1 \pm 0.3$ ps, and $\tau_3 = 0.16 \pm 0.05$ ps. The slowest relaxation process, $\tau_D$, dominates the dielectric response of the liquid water. The process originates from collective orientational dynamics of bulk water that represents the cooperative rearrangement of hydrogen bonds among water molecules, involving in their tetrahedral configuration of the hydrogen bonding network.[14, 50, 51] The next relaxation process at ~1 ps arises from weakly coordinated water molecules, charactering as the reorientation of single-water molecules, $\tau_2$.[14, 19, 50, 51, 69] The origin of the fastest relaxation process, $\tau_3$, is under discussion. For instance, Laage *et al.*,[70] and Bakker *et al.*,[71] assigned this mode to a librational reorientation of water molecules that has been phrased as hindered reorientations.[70, 76] Using a terahertz time-domain spectrometer, the fast dynamics of water at ~0.3 THz has been reported, revealing the dynamics of ions in salt solutions.[21] In a recent study of terahertz Kerr-effect in water, Elgabarty *et al.*,[10] pointed that rotational degrees of freedoms acquire energy from terahertz excitations and the energy transfer within the hydrogen bonding network takes place at ~0.5 ps. Independently, by using employing the coherent and incoherent neutron scattering technique, Arbe *et al.*,[52] reported an observation of the fastest dynamics. These results indicate that the fastest relaxation, $\tau_3$, identified in the present study based on macroscopic analysis of the dielectric response is of rotational-vibrational origin and takes place while breaking and reformation of the hydrogen bond network of water.

The dielectric response of water in aqueous salt solutions at megahertz to terahertz frequencies can be characterized by a superposition of Debye relaxation processes of water molecules. The slowest water molecules around ions form hydration layers. The orientational relaxation time of these water molecules, $\tau_{sw}$, varies from ~10 to 55 ps over a series of monovalent cations. In the aqueous environment, the dielectric response of ionic solutions also contains the above three relaxation processes of water, including the collective orientational dynamics of bulk water, $\tau_D$, at 19.25 GHz (~8.27 ps), the reorientation of single-water molecule, $\tau_2$,[14, 50, 51, 69] and structural rearrangement due to breaking and reforming hydrogen bonds, $\tau_3$.[14, 50-52, 70] The presence of ions significantly influences the dielectric strengths of all relaxation dynamics (Figs. 3 and S4). On the other hand, the relaxation times corresponding to different modes of water show only slight fluctuation with salt concentration. Zalden *et al.*,[19] showed less than 10% change in $\tau_2$ relaxation time with NaI salt solution which is also collinear with an earlier work from Tielrooij *et al.*[12, 77]

To explore the origin of the dielectric response at the terahertz frequencies, a number of experiments have been performed on ionic solutions. Using terahertz time-domain spectroscopy combined with MD simulations, Balos et al.,[21] have studied a broad range of aqueous salt solutions and have shown a relaxation dynamics occurring at ≈ 0.5 THz, and attributed it to ion's fluctuations in the solvation cage. In a separate report, Schmidt et al,[78] also using terahertz spectroscopy to investigate aqueous solutions of monovalent cations, have shown rattling motion of ions in solvation cage occurring at a slightly higher frequency than the rotational-librational mode of water at ≈ 0.15 ps. Using optical Kerr-effect spectroscopy, Heisler et al.,[79] have reported a single exponential relaxation with a relaxation time of 56 ± 8 fs and a damped-harmonic oscillation at high frequencies of 168, 150, 132 cm$^{-1}$ for NaCl, NaBr and NaI solutions, respectively. The high frequency damped-harmonic oscillation in ionic solutions appeared with a broad and well-defined peak is absent in pure water spectrum, and has been attributed to ion-water interaction. The 56 ± 8 fs single exponential relaxation have been observed in pure water and a more pronounced peak is present in salt solutions. Shalit et al.,[20] have investigated the impact of monatomic cations on the relaxation dynamics of hydrogen-bond network in aqueous salt solutions using two-dimensional Raman-terahertz spectroscopy. The results show an average relaxation dynamics intrinsic to hydrogen bond network with relaxation time of ≈ 65, 55, 80 fs for pure water, CsCl and NaCl solutions, respectively. These results indicate that this sub-picosecond relaxation mode is intrinsic to water structure and is in a reasonably good agreement with $\tau_3$ relaxation observed in the present study.

The dynamics of water molecules around cations is influenced as a result of cation-water interaction. Water molecules in the hydration shells of small cations reorient relatively slow as compared to the large ones. The dynamics of water molecules directly interacting with small cations slows down with a factor of 5 to 6 times (Table 1) compared with those of bulk water. However, the electrostatic effects of large cations such as Rb$^+$ and Cs$^+$ are constrained to the first solvation layer only, signifying local effects, with minor slowdown in the orientational dynamics of surrounding water molecules. As can be seen in Table 1, a correlation can be observed between the charge density and orientational relaxation times of water molecules having a direct contact with cations. Beyond the hydration shells, two faster relaxational dynamics observed in all aqueous salt solutions are identical to that observed in pure water. These relaxation modes are attributed to the dynamics of weakly coordinated water molecules that are largely uncorrelated form of the tetrahedral network of bulk water.

The estimation of the total number of hydration water molecules and the probability distribution of water molecules around ions, RDF$_{X-O}$(r), concurrently suggest that the electrostatic potential affects the kinetics of water molecules. The small cations with higher charge densities have greater tendency to accumulate water molecules around them. The first two layers of water molecules are strongly bound to Li$^+$. However, large cations such as K$^+$, Rb$^+$ and Cs$^+$ are weakly hydrated, and the hydration number is very low and close to zero in case of Cs$^+$ ion (Fig. 5). The hydration number decreases from ~ 8.5 to 0 from Li$^+$ to Cs$^+$, respectively.

Beyond the hydration layers of small ions (e.g, Li$^+$, Na$^+$) or the surface of large cations such as Rb$^+$ and Cs$^+$, the dynamics of water molecules is characterized by the collective reorientation, $\tau_D$, and fast relaxation processes of weakly coordinated water molecules involving in the single-water molecule reorientation, $\tau_2$, and structural rearrangement due to breaking and reforming of hydrogen bonds, $\tau_3$. The dielectric strengths of the fast relaxation processes increase with salt concentration and follow the ionic order Cs$^+$ > Rb$^+$ > K$^+$ > Na$^+$ > Li$^+$. Ions induce an increase of the fast water relaxation processes. The population density of single-water molecules and the structural rearrangement activities increases with salt concentration and follows the aforementioned ionic order. The ordering provides the effects of the ions on the water structure which is related to dynamical fluctuations of the hydrogen-bond network. Fig. 5(e) provides a schematic diagram to elaborate interfacial layers of water around Li and Cs ions along with relaxation modes of slow, fast, and bulk water relaxation processes.

The order of the ion effects on the structure of water also follows the mobility order of the monovalent cations, Cs$^+$ > Rb$^+$ > K$^+$ > Na$^+$ > Li$^+$ (Table 1).[59] In this order, the mobility is high for heavy cations, and low for light ions (Table 1). A possible explanation is that small ions such as Li$^+$ and Na$^+$ carry a large number of hydration water (i.e., they are strongly hydrated), thus, they move slowly as compared with heavy cations (e.g., Rb$^+$, Cs$^+$). Our observation of the total dielectric amplitudes of fast relaxation processes beyond the hydration shells shows that large cations such as Rb$^+$ and Cs$^+$ induce an increase in the total dielectric strengths of the fast relaxation processes, i.e., a larger volume of structure breaker in the intermediate layer. Thus, the diffusion ability of water molecules around large cations is enhanced as compared to that of small cations, explaining the order of mobility of monovalent cations in aqueous solutions.

The relation of the effects of ions on the water structure as structure making or breaking properties to the viscosity has earlier been established.[7, 80] Solvation of salts into water either enhances or diminishes the viscosity,[80] which is often explained by Jones-Dole empirical relation, $\eta/\eta_0 = 1 + Ac_s^{1/2} + Bc_s + Dc_s^2$, with $\eta$ and $\eta_0$ being viscosity of aqueous salt solution and bulk water, respectively, $c_s$ is molar concentration of salt, $A$ and $D$ depend on interatomic forces, and $B$ defines ion-water interaction. The sign of the coefficient $B$ can be negative or positive depending on the nature of interaction ($B > 0$ for structure makers and $B < 0$ structure breakers, Table 1). Such empirical classification has previously been used by Marcus *et al.*,[7, 80] and others,[17, 20, 81] to categorize the effect of ions on water as "*kosmotropic*" or "*chaotropic*". The order of cations in our observations for the hydration shells and the dielectric response of fast relaxation processes is analogous to the characterization of ions by Jones-Dole $B$ coefficients. Specially, the hydration number reduces from ~8.5 in case of Li$^+$ to 0 in case of Cs$^+$, and in opposite, the total dielectric strength of the fast relaxation processes increases from $Li^+$ to $Cs^+$. The bound water molecules interacting with the smaller cations with higher surface charge density, such as $Li^+$ and $Na^+$, are orientationally ordered under the influence of local electrostatic field of cations as compared to those weakly interacting with the large ones with lower surface charge density. This makes small cations having the least overall tendency to break intermolecular correlation among water molecules, reflecting in the dielectric strengths of fast water dynamics. One can organize the cations based on their tendency to influence the dynamic fluctuation of the hydrogen-bond network as $Cs^+ > Rb^+ > K^+ > Na^+ > Li^+$, that conforms well to the Hofmeister series for monovalent cations.

**Conclusions**

In conclusion, the presence of ions in aqueous solutions significantly affect various macroscopic properties of the solutions, thus, altering molecular activities in solution. The dielectric spectroscopy has revealed the solvation dynamics in aqueous salt solutions, which is composed of several features in the megahertz to terahertz frequency region. Monovalent cations, namely, $Li^+$, $Na^+$, $K^+$, $Rb^+$, and $Cs^+$ are chosen for the study with a common anionic chloride ion. Our study reveals that electrostatic forces of ions influence the dynamics of water in multiple processes, forming interfacial layers composed of the hydration shells and intermediate layer. In the gigahertz frequency region, the dielectric response shows that the effects of ions strongly depend on their charge densities, small cations with high charge density, such as $Li^+$ and $Na^+$, strongly influence the dynamics of water molecules, and form hydration shells around them. The large cations with low charge densities have weak electrostatic effects on the water dynamics, thus, the hydration number is low or even zero in case of $Cs^+$. In the terahertz region, the dielectric response of salt solutions is governed by the two distinct fast relaxation modes arising from weakly coordinated water molecules in the intermediate layer at the ion-water interface, including the orientation of single-water molecules and the structural rearrangement due to breaking and reforming hydrogen bonds around solvated ions. The terahertz spectroscopy reveals that this behavior is originated from an increase in the population density of weakly coordinated water, as a result of reduced orientational correlation among water molecules under the influence of cations. The tendency of these monovalent cations to break the intermolecular cooperativity among water molecules is in order, $Cs^+ > Rb^+ > K^+ > Na^+ > Li^+$.

**Acknowledgements**

This material is based upon work supported by the Air Force Office of Scientific Research under award number FA9550-18-1-0263 and National Science Foundation (CHE-1665157).

## Supplementary material

See the supplementary material for details of terahertz spectroscopy of salt solutions, molecular dynamics simulations, analysis of dielectric strength of water in salt solutions, and hydration numbers.

## Author declarations

Conflict of Interest: The authors have no conflicts to disclose.

# Interfacial Layers between Ion and Water Detected by Terahertz Spectroscopy

## Supplementary Material

Abhishek K. Singh, Luan C. Doan, Djamila Lou, Chengyuan Wen, Nguyen Q. Vinh*

Department of Physics and Center for Soft Matter and Biological Physics, Virginia Tech, Blacksburg, Virginia 24061, USA

*Corresponding author: vinh@vt.edu; phone: 540-231-3158

**Dielectric spectroscopy of salt solutions.** The dielectric response of the salt solutions has been collected in an extended frequency region from 500 MHz to 1.12 THz. The spectrometer consists of a commercial vector network analyzer (VNA) from Agilent, N5225A PNA, that covers the frequency range from 10 MHz to 50 GHz. Frequency extenders from Virginia Diodes, Inc. (Charlottesville, VA), have been used to generate terahertz waves. Eight different rectangular-waveguide (WR) modules are employed to cover a frequency range of 9 GHz – 1.12 THz as described in previous works[1]. A dielectric probe kit (HP 85070E) is interfaced with the VNA to measure the dielectric response in the frequency range from 500 MHz to 9 GHz[1-4]. A variable path length sample cell is used for the dielectric measurements of aqueous salt solutions at terahertz frequencies. Two parallel windows are mounted inside an aluminum sample cell, with once fixed and other in mobile position, in order to vary the thickness of liquid samples. A Newport (XMS160 ultra-precision linear motor stage) is employed to perform nanometer precision motion with a travel range of 160 mm. High power resistors are embedded in the aluminum sample cell in order to provide a controlled heating. The temperature is controlled with accuracy of ± 0.02 ºC, using a Lakeshore 336 temperature controller. Additional information regarding experimental setup can be accessed in earlier reports[1-9].

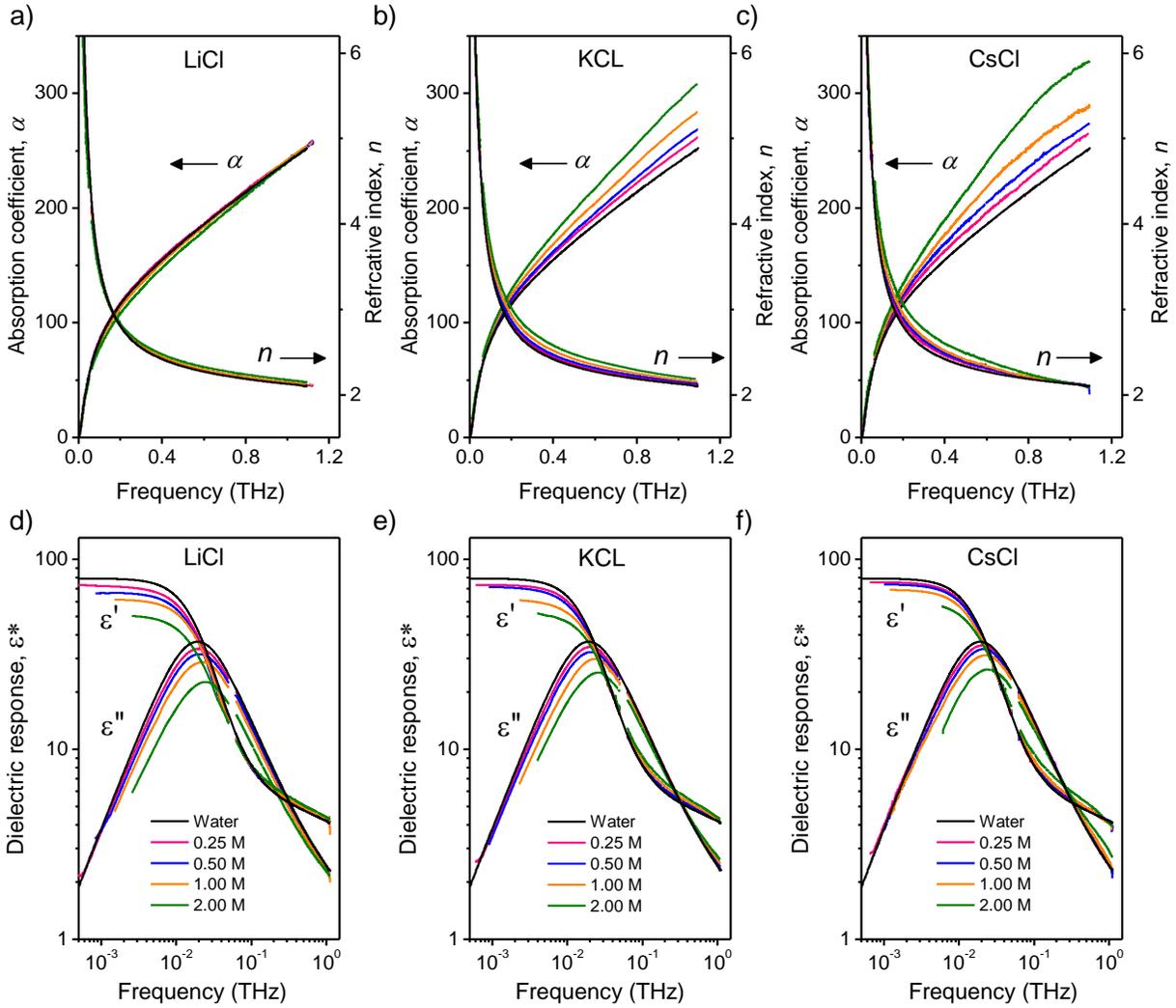

**FIG. S1.** Dielectric response of aqueous salt solutions was collected in the frequency range from 500 MHz to 1.12 THz. Plots in (a), (b) and (c) panels show the absorption coefficients, $α(v)$, and refractive indices, $n(v)$, of aqueous LiCl, KCl and RbCl solutions, respectively, at the concentrations of 0.25, 0.50, 1.0 and 2.0 M. The corresponding dielectric function, $ε^*(v)$, including the real, $ε'(v)$, and imaginary, $ε''(v)$, parts are shown in (d), (e) and (f) panels.

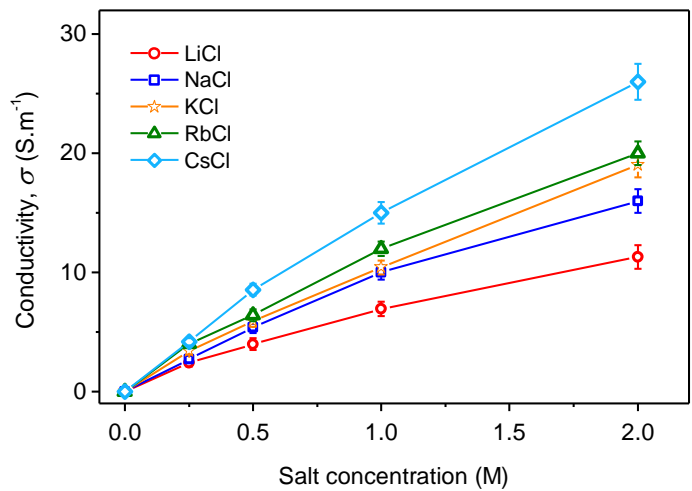

**FIG. S2.** The electrical conductivity, $\sigma$, of aqueous salt solutions including LiCl, NaCl, KCl, RbCl, and CsCl increases with increasing salt concentration.

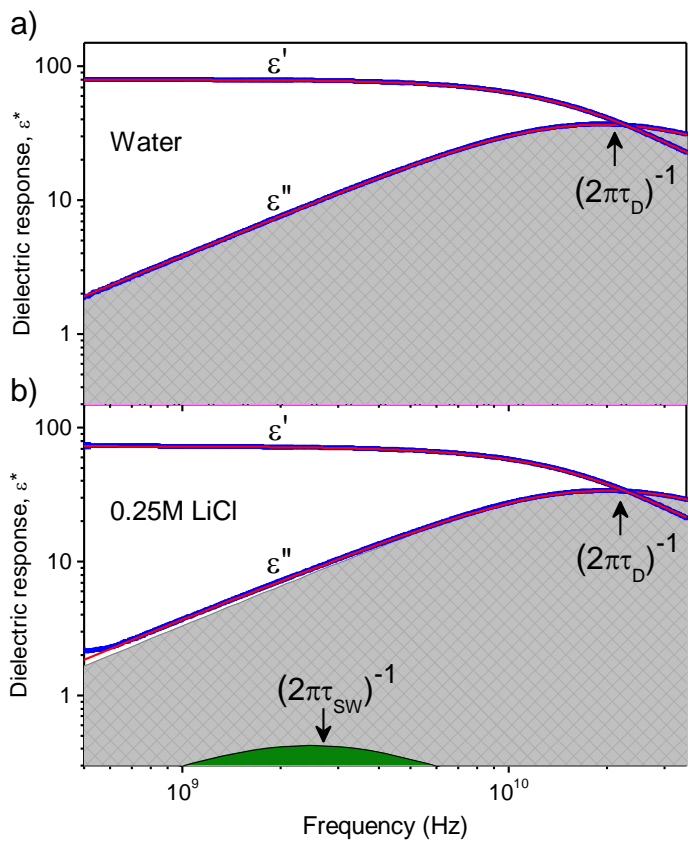

**FIG. S3.** Megahertz to gigahertz dielectric response of (a) water and (b) 0.25 M LiCl solution provides insights into the dynamics of slow water molecules in hydration shells. The red curves are fits of the real and the imaginary components of the complex dielectric response of two samples to a two-Debye model including slow water and collective reorientation water molecules.

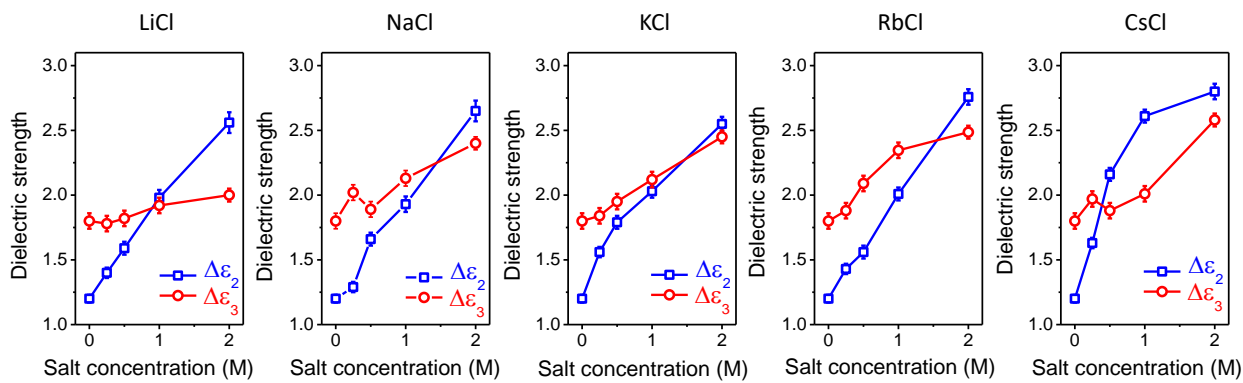

**FIG. S4a.** The dielectric amplitudes for fast dynamic processes increase significantly with increasing salt concentration. The mechanistic implications of these effects are explored in the text.

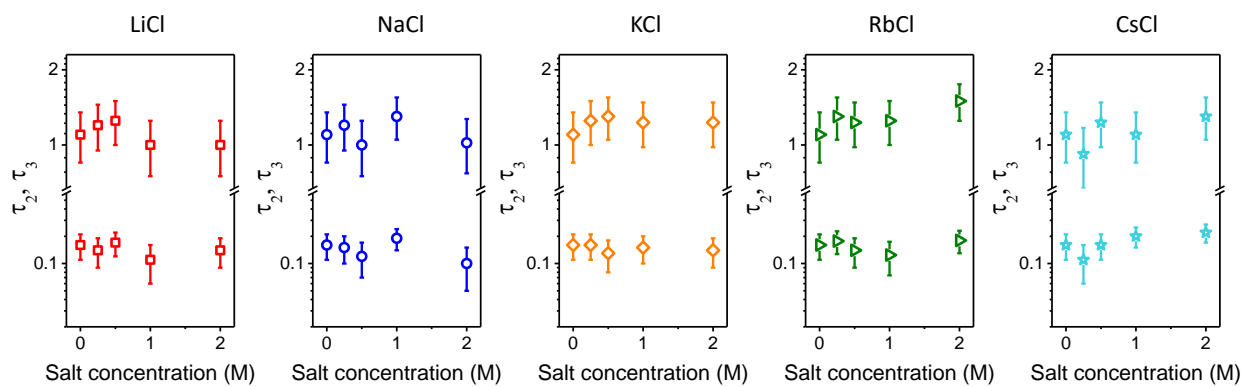

**FIG. S4b.** Fast dynamics of water dynamics are plotted as a function of salt concentration.

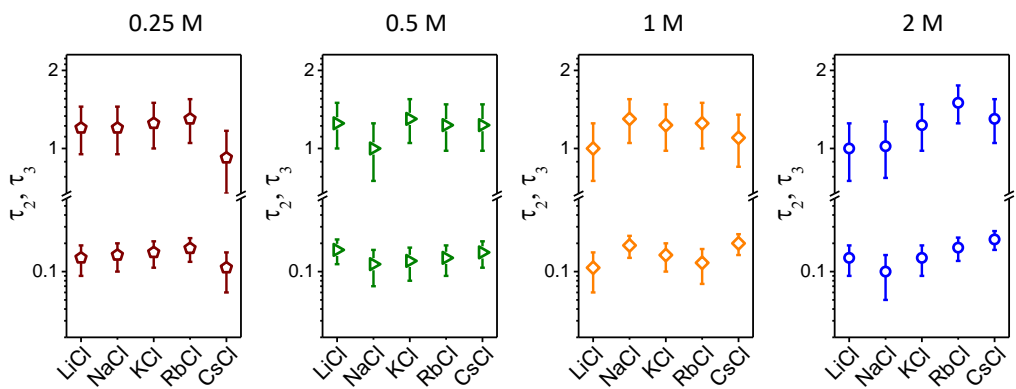

**FIG. S4c.** Fast dynamics of water dynamics are plotted as a function of the size of cations which is related to the surface charge density.

**Molecular Dynamics Simulations**. Molecular dynamics simulations are carried out to access molecular level insights and complement our experimental results of aqueous salts solutions. All the atomistic simulations are performed by means of the Large-scale Atomic/Molecular Massively Parallel Simulator (LAMMPS)[10]. SPC/E water model is used as it provides relaxation times for bulk water compatible to that obtained from the dielectric measurements. The Lennard-Jones (LJ) potential is used as the non-bonded potential in the form,

$$U(r) = 4\varepsilon \left[ \left(\frac{\sigma}{r}\right)^{12} - \left(\frac{\sigma}{r}\right)^{6} \right] \tag{S1a}$$

$$\varepsilon_{ij} = \sqrt{\varepsilon_i \varepsilon_j} \tag{S1b}$$

$$\sigma_{ij} = \sqrt{\sigma_i \sigma_j} \tag{S1c}$$

where $r$ is the distance between atoms, $\varepsilon$ is the LJ depth of potential well, and $\sigma$ is the distance at which the potential between two interacting particles becomes zero. LJ parameters of ions $Li^+$, $Na^+$, $K^+$, $Cs^+$, and $Cl^-$ are chosen from Ref. 11 and that for $Rb^+$ was chosen from the Ref. 12. The parameters used here are shown in Table S1. The geometric mix rules shown in Eq. 5b, c are used for interactions between different types of atoms. Coulombic interactions are included, and particle-particle particle mesh method is used for long-range interactions. Velocity Verlet algorithm is applied to integrate the equation of motion with a time-step of 1 fs. A total of 6097 water molecules in a cubic box are built by NAMD tool. The periodic boundary conditions are applied in x, y, z directions. A total of 72 cation-anion pairs are distributed to form 0.5 M of the salt solutions. The systems are equilibrated at 298 K and 1 atmosphere for 5 ns in an NPT ensemble with Nose Hoover algorithm. After equilibration, another 5 ns simulation at 298 K with NVT ensemble using Nose Hoover algorithm is taken for the production run.

**Table S1**. The LJ parameters of salt ions and water.

| Atom | $\varepsilon$ (kJ/M) | $\sigma$ (Å) |
|---|---|---|
| H | 0.0 | 0.0 |
| O | 0.650 | 3.17 |
| Cl | 0.419 | 4.4 |
| Li | 0.69 | 1.506 |
| Na | 0.544 | 2.35 |
| K | 0.568 | 3.36 |
| Rb | 0.419 | 3.53 |

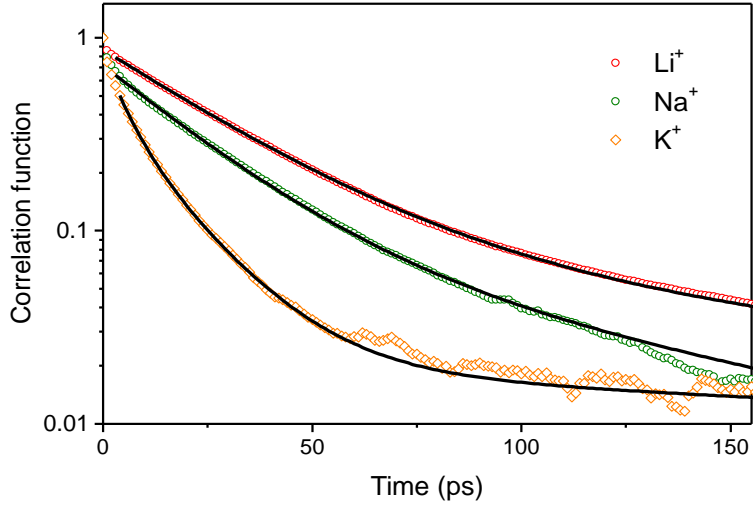

**FIG. S5.** Fitting the cation-water orientational correlation functions, $C(t)_{X-O}$ (X stands for monovalent cation), to Eq. 2 in the main text is performed for monovalent cations. The relaxation times are provided in the main text (Table 1).

The orientational autocorrelation function is calculated using,

$$C(t) = \langle \frac{1}{N} \sum_{i=1}^{N} \mu_i(t)\,\mu_i(0) \rangle \tag{S2}$$

where $\mu_i(t)$ is the unit vector in the direction of the instantaneous electric dipole associated with a water molecule at the time $t$, and the summation is over the total $N$ water molecules in the aqueous solution. The ensemble average is computed under the assumption of different states of the system as an initial state at $t = 0$. Our analyses indicate that orientational autocorrelation functions, $C(t)$, can be fitted to a superposition of three exponential functions, $\sum_i A_i e^{-(t/\tau_i)}$, where $\tau_i$ is the orientational relaxation time corresponding to $i^{th}$ relaxation mode, and $A_i$ being a weighted coefficient for that particular mode. The $A_i$ and $\tau_i$ parameters obtained from the fitting are shown in Table 1 (we use $\tau_A$, $\tau_B$, and $\tau_C$ for $\tau_1$, $\tau_2$, and $\tau_3$). Our MD simulations for pure water show that the orientational relaxation time of water molecules is $5.2 \pm 0.6$ ps[3, 4, 13, 14], which is in good agreement with other reports of 4.5 ps[15, 16].

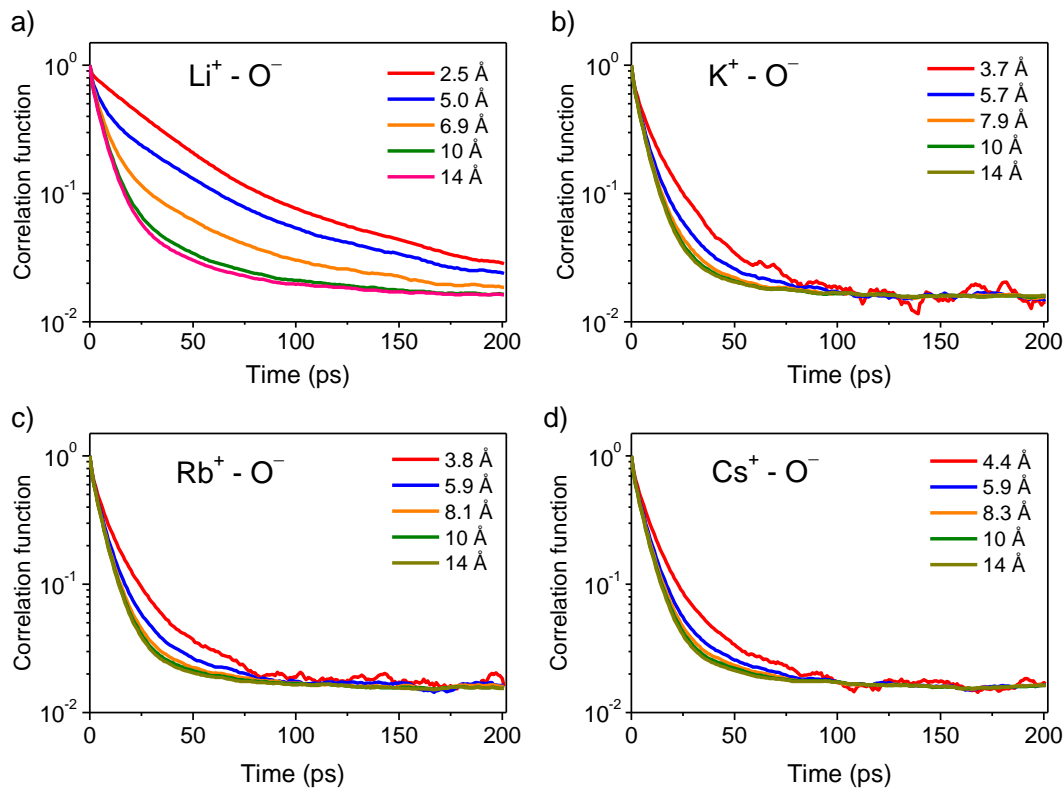

**FIG. S6.** The orientation correlation functions of water around cations ($X^+ - O$, where X stands for cation, and O is the oxygen of water molecules) as a function of the distance from the cation for $Li^+$, $K^+$, $Rb^+$, and $Cs^+$ until 14 Å.

**Table S2:** The dynamics of water molecules in the first solvation shell extracted from orientational correlation functions. The orientational relaxation times ($\tau_A$, $\tau_B$, and $\tau_C$) have been obtained from fitting of orientational correlation functions, $C(t)$, to a three-exponential component function. For comparison, relaxation times from literature are included.

| Ions | First minimum (Å) | Second minimum (Å) | $\tau_A$ ± 1 (ps) | $\tau_B$ ± 2 (ps) | $\tau_C$ ± 30 (ps) |
| --- | --- | --- | --- | --- | --- |
| $Li^+$ | 2.5 | 5.0 | -- | 32.0 (50.6[16]) | 180 (220[17]) |
| $Na^+$ | 3.1 | 5.4 | -- | 25.0 (22.4[16], 26.4[15]) | 160 (203[18]) |
| $K^+$ | 3.7 | 5.7 | 5.2 | 12.0 (14.3[16], 9.4[15]) | 150 |
| $Rb^+$ | 3.8 | 5.9 | 5.1 | 11.5 (12.1[16], 9.3[15]) | 150 |
| $Cs^+$ | 4.4 | 5.9 | 5.9 | 10.5 (13.9[16], 9.5[15]) | 140 (143[19]) |
| $Cl^-$ | 3.8 | 6.1 | 5.0 | 8.8 (9.0[15]) | -- |

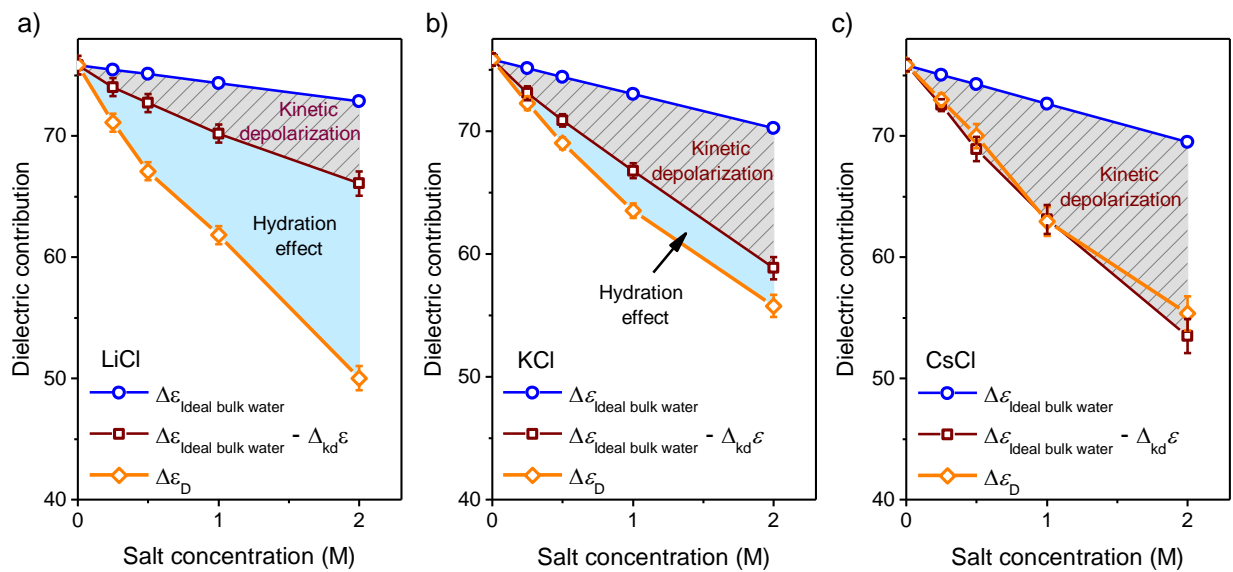

**FIG. S7.** The dielectric strength of bulk water in salt solutions including LiCl, KCl, and CsCl decreases with increasing salt concentration. The blue line represents the dielectric amplitude of ideal bulk water from analysis of water concentration in the solutions under an assumption that all water molecules in the solutions contribute to the collective orientational dynamics of bulk water. A correction for the kinetic depolarization contribution originated from moving ions in an electrical field is provided. The light-blue area is the contribution of slow water molecules in the hydration shell (i.e., hydration effect) in the solution.

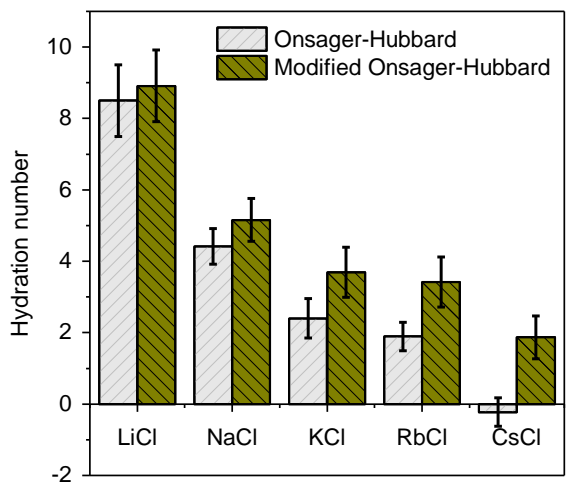

**FIG. S8**. A comparison of hydration numbers predicted using the Onsager- Hubbard model and phenomenological modification of the Onsager- Hubbard model at 0.25 M salt solutions.

**Hydration number**. Hydration numbers estimated using the Onsager – Hubbard continuum theory[20] for all salt solutions at concentration of 0.25 M (Eq. 4a,b) are shown in Fig. 5d. We also employ a phenomenological modification of the Hubbard–Onsager continuum theory proposed by Sega *et al.*,[21] which consider the screening due to the ionic cloud at the mean-field level for the kinetic depolarization,

$$\Delta_{\text{kd}}\varepsilon(c_s) = \frac{2}{3}\left(\frac{\varepsilon_s(0) - \varepsilon_\infty(c_s)}{\varepsilon_s(0)}\right)\frac{\tau_D(0)}{\varepsilon_0}\sigma(c_s)\exp(-\kappa R), \qquad (S3)$$

where $\kappa$ is the inverse Debye screening length and $R$ is the effective ionic radius. A comparison is shown in supplementary material, Fig. S8. As shown the correction affects hydration numbers more for bigger cations than the smaller ones.